\documentclass[12pt]{article}
\usepackage{a4wide}
\usepackage{graphics}
\usepackage{graphicx}
\usepackage{amsmath}

\newcommand{\Xp}{\Xi_{cc}^+}
\newcommand{\Xpp}{\Xi_{cc}^{++}}
\newcommand{\Lc}{\Lambda_c^+}
\newcommand{\Jp}{J/\psi}

\newcommand{\MeV}{\mbox{\,MeV}}
\newcommand{\GeV}{\mbox{\,GeV}}

\newcommand{\fm}{\mbox{\,fm}}

\begin{document}

\thispagestyle{empty} 

\begin{center} 
{\bf\large
Analysis of the baryonic state $|[qc]c\rangle$}\\[1.0cm]
{\large Sergey~Koshkarev and Stefan~Groote}\\[0.3cm]
Institute of Physics, University of Tartu, 50411 Tartu, Estonia\\[3pt]
\end{center}

\vspace{0.2cm}
\begin{abstract}
In this note we analyse the structure of the baryonic state $|[qc]c\rangle$
with a spin-0 diquark $[qc]$, as compared to baryonic states $|(qc)c\rangle$
and $|q(cc)\rangle$ containing internal spin-1 diquark states. While one can
identify the state $|q(cc)\rangle$ with the state observed at LHCb, the state
$|(qc)c\rangle$ could probably be part of the state $\Xi_{cc}^{++}(3780)$.
Accordingly, the ground state of $|[qc]c\rangle$ can be identified with the
state $\Xi_{cc}^+(3520)$.
\end{abstract}

\section{Introduction}

Since the publication of first evidences for the existence of doubly charmed
baryons by the SELEX collaboration, this result became probably the most
intriguing and  controversial one in modern baryonic
physics~\cite{Mattson:2002vu,Ocherashvili:2004hi}. The reason for this is that
perturbative QCD can explain neither the SELEX production rate nor the $x_F$
distribution.

However, during the past two years significant progress has been made in
understanding the SELEX result. By involving the intrinsic charm mechanism it
was possible to explain the production rate and the kinematics
dependencies~\cite{Koshkarev:2016rci,Groote:2017szb}.

The most recent LHCb result on the production of the doubly charmed
baryons~\cite{Aaij:2017ueg} illuminated the significant gap between the
$3520\MeV/c^2$ SELEX event and the mass $3621.40\pm 0.72\text{(stat)} \pm
0.27\text{(sys)}\pm 0.14(\Lc)\MeV/c^2$ measured by LHCb. This apparent conflict
can be solved by introducing the state $|[qc]c\rangle$ containing a spin-0
diquark $[qc]$, as this state can be naturally produced by the charmed Fock
state comoving with the same rapidity as the valence quarks of the projectile
hadron~\cite{Brodsky:2017ntu}. In this note we briefly review some
consequences of the structure of the baryonic state $|[qc]c\rangle$. 

\section{Isopin splitting of the SELEX states}
The analysis of the isospin splitting of the SELEX states implies that double
charm  baryons are very compact, i.e.\ the light quark must be very close to
the two heavy quarks~\cite{Brodsky:2011zs}. This contradicts the usual wisdom.
Indeed, within the heavy-diquark concept, the production of the doubly charmed
baryon can proceed in two steps. In a first step, due to the reactions
$q\bar q\to c\bar cc\bar c$ or $gg \to c\bar cc\bar c$ the production of two
$c$ quarks with a small relative momentum will take place, followed by the
formation of a $cc$-diquark in the color-antitriplet state. In a second step,
the transition of the produced diquark into the baryon is performed. The
normalization of the fragmentation of the $cc$-diquark into the double-charm
baryons is unknown. However, one is still able to provide some quantitative
analysis because the fragmentation function is proportional to the wave
function at the origin. The color-antitriplet wave function can be estimated
on the basis of information about the color-singlet wave function,
$|R(0)[cc]_{\bf\bar 3}|\sim|R(0)[c\bar c]_{\bf 1}|$. This leads to an atom-like
structure where the $cc$ diquark forms the compact core while the scale of the
light quark is given by the nonperturbative confinement
scale~\cite{Kiselev:2001fw}. In case of the $S$-wave solution we have the
scale hierarchy
\begin{equation*}
r_{cc} : r_{QCD} \approx 0.39 : 1 \,
\end{equation*}
where $r_{cc}\sim r_{\Jp}\sim 0.39\fm\simeq(0.5\GeV)^{-1}$~\cite{Nochi:2016wqg}
and $r_{QCD}=1/(\Lambda_{QCD}\approx 200\MeV) \approx 1\fm$. 

In contrast to this, the production of the doubly charmed baryons at the SELEX 
experiment is supposed to be due to re-coalesce from a higher Fock state of the
proton such as~\cite{Koshkarev:2016rci,Brodsky:2017ntu}
\begin{equation*}
|[uu]_{\bf\bar 3}[dc]_{\bf\bar 3}c_{\bf 3}[\bar c \bar c]_{\bf 3}\rangle
\end{equation*}
which leads to the baryon state $|[qc]c\rangle$ in a natural way. Here the
scale will be characterized by the size of the spin-0 $[qc]$ diquark, given by
the Compton wavelength $\lambda_{[qc]}\sim 1/m_{[qc]}$ of the diquark. This
naturally provides closeness of the light quark to the two heavy quarks. Note
that the peculiarities mentioned here are due to the inclusion of two heavy
quarks.

It is interesting to estimate the compactness of such state. The
mass $m_{[qc]}$ can be estimated as the effective diquark mass,
$m_{\Xi_{cc}}-m_{c}$, where $m_{\Xi_{cc}}$ is the doubly charmed baryon mass
and $m_{c}$ is the mass of the $c$ quark. In case of the SELEX $3520\MeV$
event, one has $\lambda_{[dc]} \sim 0.5\fm$ which is again in the perfect
agreement with the compactness of the SELEX state calculated from isospin
splitting~\cite{Brodsky:2011zs}.

As is emphasized e.g.\ in Ref.~\cite{Chen:2017sbg}, the Fierz identity holds
only in case of local field operators. In this case the states $|[qc]c\rangle$
and $|q(cc)\rangle$ would be the same state. However, one might think about a
nonlocal baryon field operator. Based on works of Diakonov and
Petrov~\cite{Diakonov:1983hh,Diakonov:1987ty} and examplified in the instanton
liquid model~\cite{Anikin:2000rq}, a nonlocal extension of the
Nambu-Jona-Lasinio (NJL) model is proposed~\cite{Plant:1997jr,Bowler:1994ir}
which, besides being renormalizable and providing confinement for the quarks
(in contrast to the original (local) NJL model, see e.g.\
Refs.~\cite{Vogl:1991qt,Klevansky:1992qe,Bijnens:1995ww}), results in a
compact baryon (see e.g.\ Sec.~6.1 of Ref.~\cite{Rezaeian:2005ry}).

\section{The $\Xpp(3780)$ state}

At a few conferences~\cite{Moinester:2002uw,Engelfried:2007at} (cf.\ also the
PhD thesis of Mark E.~Mattson~\cite{Mattson:2002dz}), the SELEX collaboration
presented a decay process $\Xpp(3780) \to \Lc K^- \pi^+ \pi^+$ for the state
$\Xpp(3780)$ with statistical significance of $6.3\,\sigma$. By removing the
slower part of the $\pi^+$'s, SELEX observed that roughly $50\%$ of the signal
events above background decay weakly and $50\%$ decay strongly (to $\pi^+\Xp$).
However, this is not possible for a single state. As SELEX did not find a
plausible explanation for its decay properties, the result was not published.

Assuming $\Xpp(3780)$ to be an excited state $|ucc^*\rangle$, we predict the
mass of $|ucc^*\rangle$ by utilizing the predictions of supersymmetric light
front holographic QCD (SUSY LFHQCD). This approach was developed by imposing
the constraints from the superconformal algebraic structure on LFHQCD for
massless quarks~\cite{Dosch:2015nwa}. As has been shown in
Refs.~\cite{Dosch:2015nwa,Dosch:2015bca}, supersymmetry holds to a good
approximation, even if conformal symmetry is strongly broken by the heavy
quark mass.

Let us remind the reader that supersymmetric light front holographic QCD, if
extended to the case of two heavy quarks, predicts that the mass of the
spin-1/2 baryon should be the same as the mass of $h_c(1P)(3525)$
meson~\cite{Dosch:2015bca}. This is well compatible with the SELEX measurement
of $3520.2\pm 0.7\MeV/c^2$ for the $\Xp$.

The SUSY LFHQCD prediction for the baryon mass spectra is given by the simple
formula
\begin{equation*}
M^2 \propto \lambda (n + L + 1) \,
\end{equation*}
where $\sqrt{\lambda}\approx 0.52\GeV$ is the fundamental mass parameter, given
by the characteristic mass scale of QCD~\cite{Brodsky:2016yod}. Using this
simple formula, we can estimate the masses of states $|[qc]c\rangle_{3/2}$
($n=1$, $L=0$) and $|(qc)c\rangle_{3/2}$ ($n=0$, $L=1$), where $(qc)$ indicates
the spin-1 diquark. These states should have the same mass around
$3730\MeV/c^2$, where the uncertainty of SUSY LFHQCD predictions is at least
of the order of $100\MeV$. Obviously, we have good agreement with the data for
$\Xpp(3780)$.

Investigating the decay properties, the $|[qc]c\rangle_{3/2}$ is more
preferable for the weak decay. In contrast to that, $|(qc)c\rangle_{3/2}$
includes a $D^*$-meson-like state, leading to the strong decay
$(qc)\to[qc]+\pi$, similar to $D^* \to D + \pi$.

As shown in Ref.~\cite{Wilczek:2004im, Selem:2006nd}, such a structure does
not lead to additional states for baryons consisting only of light quarks
($m_{u,d,s}<\Lambda_{\rm QCD}$). Additional states are found neither for
baryons containing both the $c$ and the $b$ quark, as the structure $[bq]c$
does not affect the scale hierarchy~\cite{Kiselev:2001fw}
\begin{equation*}
\lambda_b : \lambda_c : r_{bc} : r_{QCD} \approx 1 : 3 : 9 : 27 .
\end{equation*}
This fact leads to the same ground state masses: the prediction
$M(|[bq]c\rangle)=6750\pm 100\MeV$ of SUSY LFHQCD~\cite{Nielsen:2018ytt},
deduced from $M(|[bq]c\rangle)\sim M(B_{c1})$ is consistent with the
prediction $M(\Xi_{bc})\approx 6820\MeV$ of the potential
model~\cite{Kiselev:2001fw}.

\section{Summary}

In this note we investigated some consequences of the baryonic state
$|[qc]c\rangle$. As shown in Ref.~\cite{Brodsky:2017ntu}, this state can solve
the apparent SELEX--LHCb doubly charmed baryon conflict. The theoretical
estimate for the compactness of the baryon is in good agreement with similar
estimates from isospin splitting. In addition, we gave estimates for the mass
and the decay properties of $\Xpp(3780) \to \Lc K^- \pi^+ \pi^+$.

\subsection*{Acknowledgements}
This research was supported by the Estonian Research Council under Grant
No.~IUT2-27.

\end{document}